# NMR study on the local structure in La-based high-$T_c$ cuprates

**T Goto[1,3], J Tonishi[1], T Suzuki[2,1], A Oosawa[1], M Fujita[3], K Yamada[3], T Adachi[4] and Y Koike[4]**
[1] Faculty of Science and Technology, Sophia University, Kioicho, Chiyodaku, Tokyo 102-8554, Japan
[2] Advanced Meson Science Laboratory, Nishina Center, RIKEN, Wako, Saitama 351-0198, Japan
[3] Institute for Materials Research, Tohoku University, Katahira, Sendai 980-8577, Japan
[4] Graduate School of Engineering, Tohoku University, Aramaki, Aobaku, Sendai 980-8579 Japan

E-mail: gotoo-t@sophia.ac.jp

**Abstract.** To date, there has been no evidence for the macroscopic structural phase transition to the low temperature tetragonal structure (LTT) with a space group *P4₂/ncm* in high-$T_C$ cuprate of rare earth-free $La_{2-x}Sr_xCuO_4$ (LSCO). By investigating Cu-NMR on single crystals, we have found that spatially incoherent LTT structure emerges below 50 K in the sample with $x$=0.12. This incoherent structure is considered to play a key role for the slight depression of the superconductivity around $x$=1/8.

## 1. Introduction

La-based high-$T_c$ cuprates shows an anomalous suppression of $T_c$ at the hole carrier concentration at around $x$ =1/8, where the one dimensional segregation of holes and spins known as the stripe order emerges[1,2,3]. Though the dynamically fluctuating stripe is believed to be one of the candidate for the mechanism of the high-$T_c$ superconductivity, it destructs the superconductivity when stabilized by the matching effect of the underlying lattice structure, that is, the low temperature tetragonal (LTT) phase with the space group symmetry $P4_2/ncm$ or the low temperature less orthorhombic phase (LTLO) with *Pccn*.[4,5]

It has been proved by neutron experiments that the charge stripe is stabilized and the superconductivity is strongly suppressed only when LTT or LTLO structure appears at low temperatures as in the case of $La_{2-x}Ba_xCuO_4$ (LBCO) and Nd or Eu-doped $La_{2-x}Sr_xCuO_4$ (LSCO). On the other hand, the structure of the rare earth-free LSCO remains to be low temperature orthogonal (LTO) with the space group symmetry *Cmca* at low temperatures, though the significant dip in the $x$-$T_c$ curve does exist at $x \approx 1/8$. Though the fluctuation of the structural phase transition to LTT have been repeatedly reported for LSCO with $x \approx 1/8$ by experiments of elastic constant, X-ray Absorption Fine Structure (XAFS) and inelastic neutron scattering[6,7,8], there has been reported no evidence for the macroscopic phase transition to LTT phase.

---

[1] gotoo-t@sophia.ac.jp

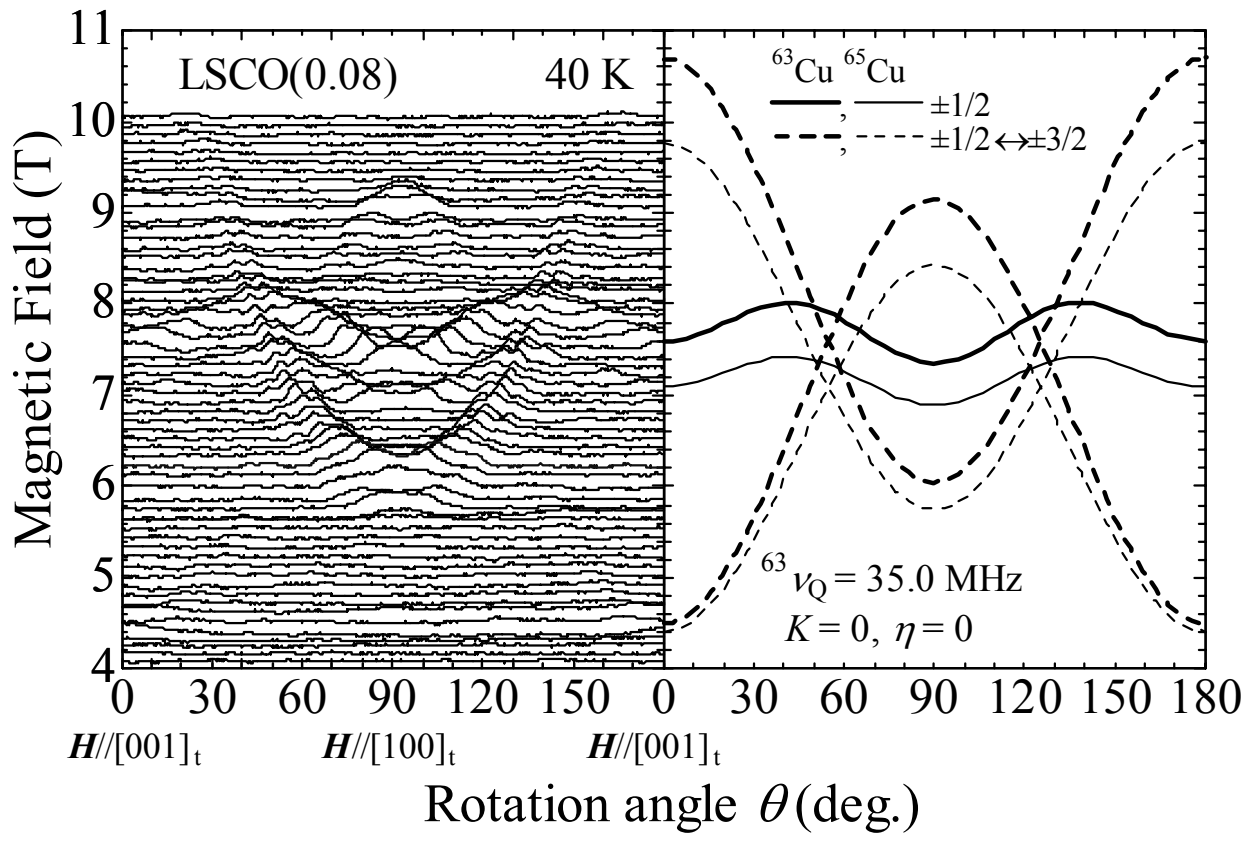

**Figure 1.** (Left) Typical rotational spectra of $^{63/65}$Cu-NMR in LSCO ($x$=0.08) [18]. They are taken under constant fields 4-10 T. (Right) Theoretical curves of each transitions calculated by the second order perturbation theory.

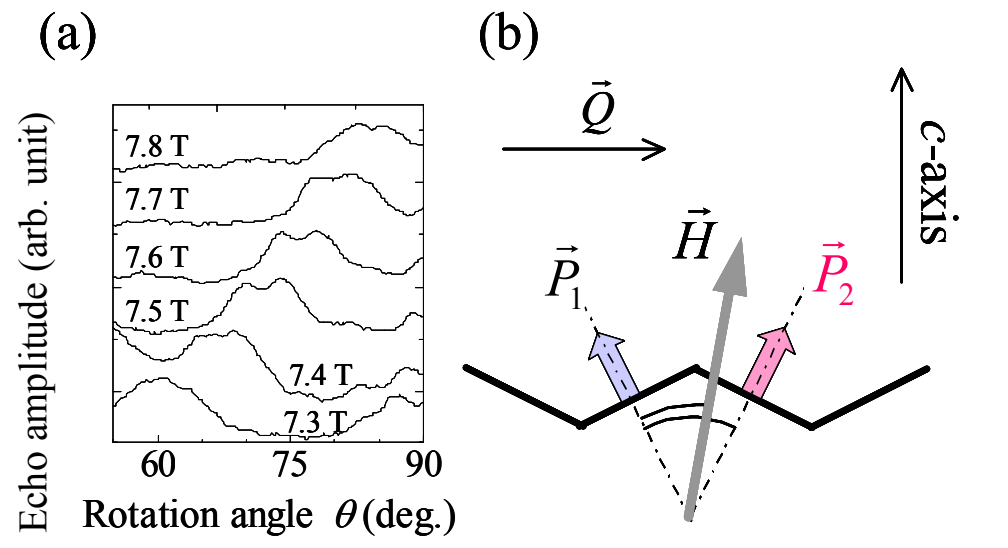

**Figure 2.** (a) Typical splitting of NMR resonance peaks due to the buckling of $CuO_2$ plane. (b) Principle of the detection of the buckling by NMR. $\vec{P}_1$ and $\vec{P}_2$ are the principal axis of the electric field gradient tensor for the two neighbouring Cu sites, $\vec{H}$ is the applied field, and $\vec{Q}$ is the buckling wave vector.

By utilizing NMR technique on high-quality single crystals, we have directly observed the buckling pattern in $CuO_2$ plane, which changes with the local structure. In this article, after a brief description of how to detect local structure by NMR, we state our first observation of static and spatially incoherent LTT structure in rare earth-free LSCO ($x$=0.12). We also refer to the observation of the incoherent LTO structure [9-14] in the vicinity of the structural phase transition from the high temperature tetragonal (HTT) with the space group *I4/mmm* to LTO in LSCO ($x$=0.15).

## 2. Experimental

The single crystals of LSCO with $x$=0.12 ($T_C$=25 K) and $x$=0.15 ($T_C$=37 K) were grown by the travelling solvent floating zone (TSFZ) method [5]. We did not detwin any of the crystals. The superconducting transition temperature is determined as the onset of a diamagnetic signal under an applied field of 20 Oe. The structural phase transition temperature from HTT to LTO in the sample with $x$=0.12 and 0.15 is determined to be 240 and 180 K respectively from the temperature dependence of the ultrasonic sound velocity as described in Appendix [6,15]. Cu-NMR "angle-swept spectra" were obtained by plotting the spin-echo amplitude against the sample rotation angle, with keeping magnetic field and resonance frequency to be constant [17,18]. Typical angle-swept spectra are shown in Fig. 1 [18]. NMR detects the buckling in $CuO_2$ plane through the electric-quadruple interaction contained in the nuclear spin hamiltonian, which is expressed as

$$\mathcal{H} = -\gamma\hbar\vec{H}_0 \cdot \vec{I} + \tfrac{1}{2}\nu_Q I_z^2 \qquad (1)$$

where $\gamma$ and $\nu_Q$ are the gyromagnetic ratio and the quadruple frequency of $^{63}$Cu nucleus, respectively, and $\vec{H}_0$ is the applied field. Note that $z$-axis is chosen to be along the principal axis of the electric field gradient (EFG) tensor, which is assumed to be perpendicular to the basal plane of $CuO_6$ octahedron, as shown in Fig. 2. Here we ignore the Knight shift and the asymmetry parameter $\eta$, since the effect of the former is much smaller than that of $\nu_Q$ and $\eta$ is considered to be very small due to the local symmetry around the Cu site. Curves in the left panel of Fig. 1. which are calculated with the second order perturbation theory based on this Hamiltonian, well reproduce the observed spectra.

As shown in Fig. 2, when the field is tilted from the $c$-axis toward the direction of the buckling wave vector $\vec{Q}$, the neighbouring two nuclei become inequivalent, so that there appears a small split in each peak in the spectrum. However, the peak does not split when the tilting direction is perpendicular to $\vec{Q}$. For example, when the magnetic field is tilted toward $(100)_t$ direction, the peak splits in LTT, but does not in LTO. This difference persists even in twinned crystals, and hence always allows one to determine the local structure from the observed pattern of split peaks. Furthermore, NMR can detect local structures irrespective of with or without the translational symmetry, because the nucleus is a point probe. Note here that in this paper the crystal axis is defined in the notation of tetragonal phase HTT.

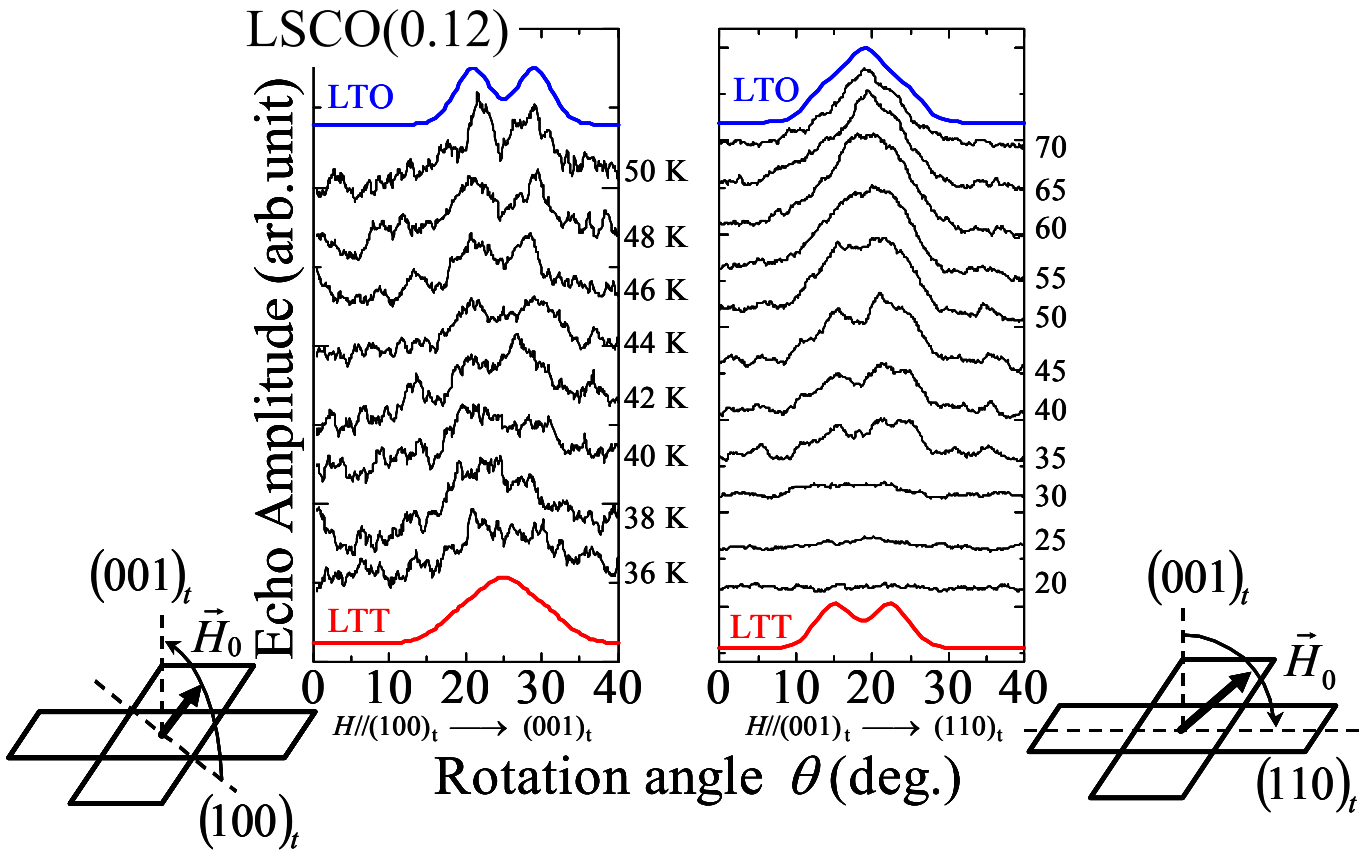

**Figure 3.** Rotational spectra of $^{63}$Cu-NMR in LSCO ($x$=0.12). Observed peaks correspond to the satellite transition of $^{63}$Cu. In the left panel, the magnetic field of 15 T is rotated within $(010)_t$ plane, while in the right, the field of 15.6T is rotated within $(1\underline{1}0)_t$ plane. Curves at the top and the bottom are calculation based on the assumption of the local structure to be LTO or LTT.

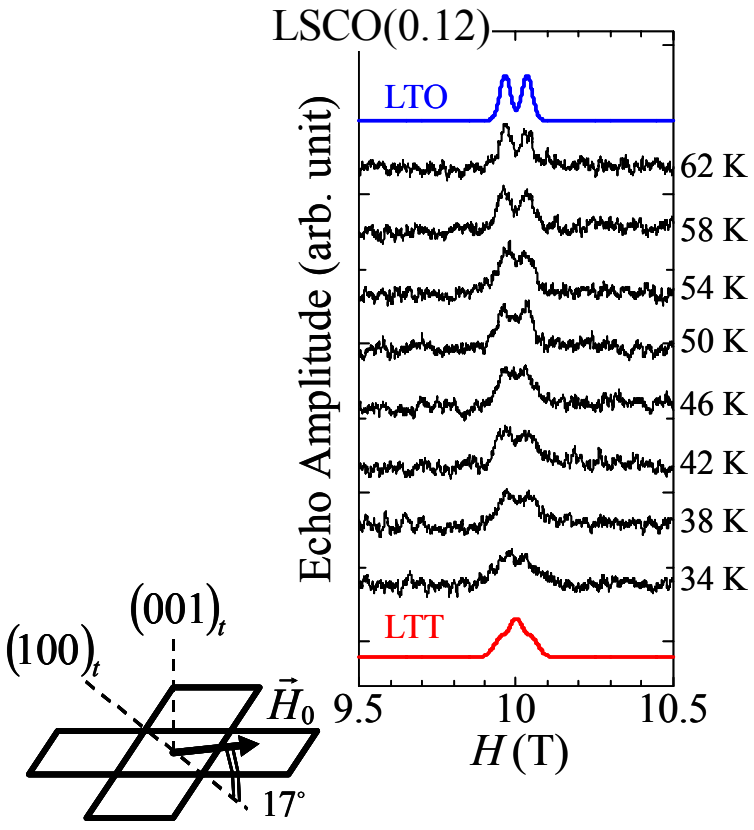

**Figure 4.** Field-swept NMR spectra for LSCO ($x$=0.12). The peak corresponds to the central transition of $^{63}$Cu. The direction of the magnetic field is along $(\cos 17°, 0, \sin 17°)_t$. Curves are the calculation based on the assumption of the local structure to be LTO or LTT.

## 3. Results and Discussion

Figure 3 shows rotational spectra for the sample $x$=0.12 taken with the magnetic field tilted from $c$-axis to $(100)_t$ or $(110)_t$. In either direction, the peak profile changes gradually from 50 K and the entire signal disappears below 30 K. We simulated curves to fit the observed rotational spectra in following steps. First, we assumed a set of EFG principal axes corresponding to the buckling pattern of $CuO_2$ planes, and calculate the resonance positions. In LTO phase with twin domains, there are four directions of EFG principal axes $(\pm\sin\theta_P, 0, \cos\theta_P)_t$ , $(0, \pm\sin\theta_P ,\cos\theta_P)_t$, where $\theta_P$ is the tilting angle of the basal plane of $CuO_6$ octahedron measured from $c$-axis. Next, those resonance positions are convoluted with a gauss function with a width of $\Delta\theta$ to obtain calculated curves shown in Fig. 3. The values of $\theta_P$ and $\Delta\theta$ were determined by data fitting. The former $\theta_P$=3.6° stays constant in the measured temperature range within the experimental precision. The gaussian width $\Delta\theta$, which is proportional to the observed peak width, showed slightly increase with decreasing temperature from 2.5° at 170 K to 3.1° at 35K.

At high temperatures above 50 K, the observed spectrum agrees well with the calculation based on the LTO-type buckling. The spectrum starts to change its profile at around 50 K, and at 40 K, the spectrum agrees with the calculation based LTT-type buckling, indicating an emergence of LTT structure. The same change in the structure below 50 K is also confirmed in the field-swept spectra shown in Fig. 4, where the split peaks unite into single one at low temperatures, indicating the change in the structure to LTT. Buckling angle $\theta_P$ determined from these spectra agrees with that from rotation spectra. This LTT structure must be spatially incoherent[8,17], because no evidence for a macroscopic structural transition to the LTT phase has so far been reported in this system. This spatially-incoherent structure is considered to be static in time, because NMR spectra only see the static components of EFG or hyperfine field.

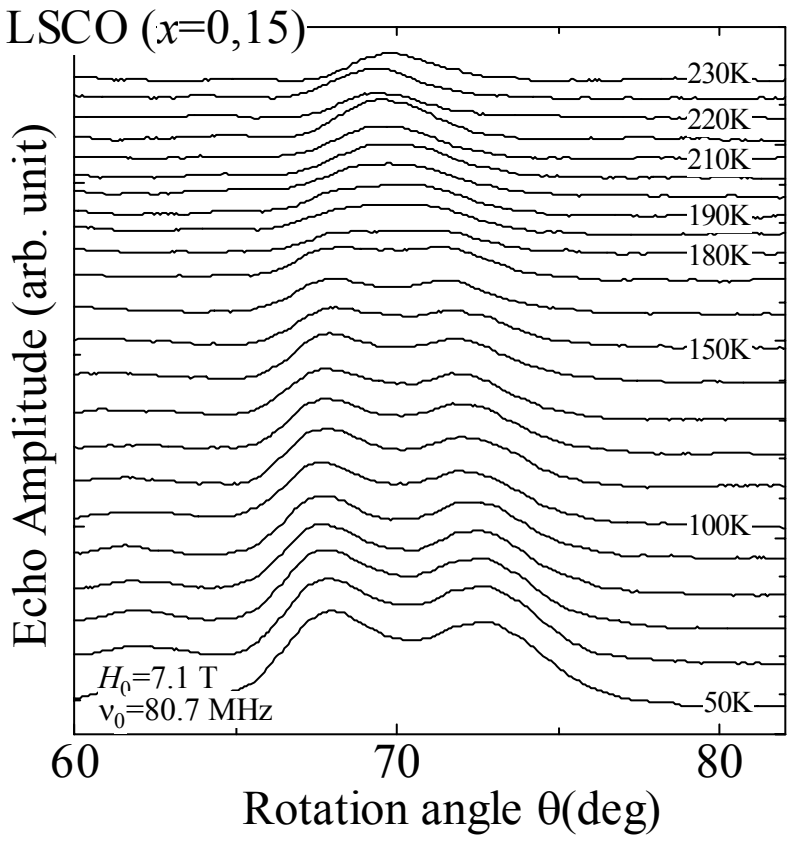

**Figure 5**. Rotational spectra of the $^{63}$Cu central transition in LSCO (0.15) at various temperatures. The field direction is rotated within (sin29°, –cos29°, 0)-plane [17].

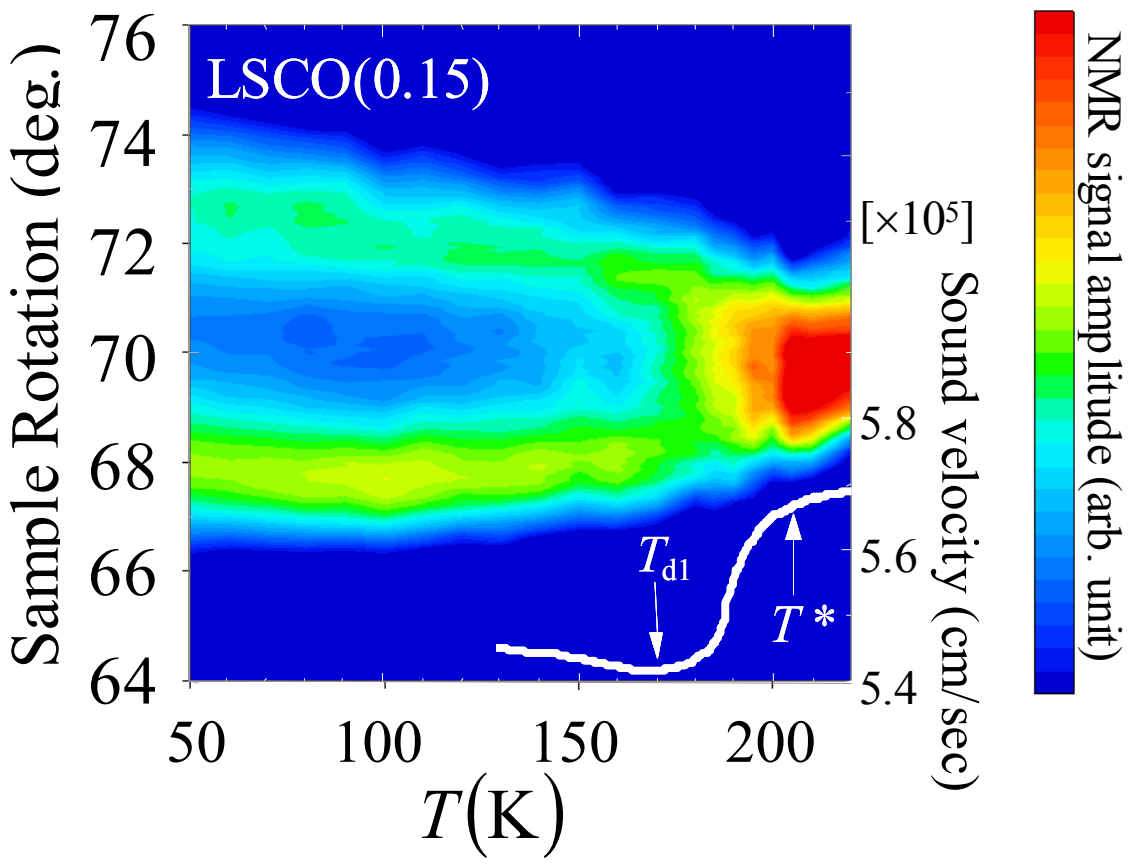

**Figure 6.** Contour map of rotational spectra in various temperatures for LSCO ($x$=0.15). A curve is the temperature dependence of the sound velocity of $c_{11}$ mode [17].

The tilting angle stays constant in the measured temperature region irrespective of the structure of LTO or LTT, that is, only the tilting direction changes on the emergence of incoherent LTT structure. This is the same as the case of coherent and macroscopic LTT phase as is reported in LBCO or Nd-doped LSCO [19,20]. From these results, we reach the first conclusion that there exists static LTT structure in rare-earth free LSCO with $x$=0.12 below 50 K, and that it is spatially incoherent. As is reported by neutron experiments [5,16], only the spin density wave (SDW) is stabilized is LSCO, and hence it is considered that the coherent LTT phase is indispensable for the occurrence of the charge density wave (CDW). Therefore, we can further conclude that this incoherent LTT structure stabilizes only SDW but not CDW. This *odd* order must set the $T_c$ anomaly at $x$=0.12 very weak in the rare earth free-LSCO [5,6]. Consequently, the disappearance of signal below 30 K is considered to be due to the wipe out effect of the SDW.

The origin of the peak width $\Delta\theta$ is considered to be the distribution of the tilting direction of $CuO_6$ octrahedron. A moderate distribution of the Knight shift and $\nu_Q$, that is, 0.2 % and 1 MHz, respectively, are ruled out for the origin, because they cannot produce the observed large value of $\Delta\theta$. The increase in $\Delta\theta$ in low temperatures can be reasonably understood if one notes that the phase inversion of the buckling occur quite frequently in the incoherent LTT structure.

Next, we show in Fig. 5 rotational spectra of LSCO ($x$=0.15) in the vicinity of the structural phase transition from HTT to LTO phase. At high temperatures in HTT phase with flat $CuO_2$ plane, all the Cu sites are equivalent and single peak is observed. With decreasing temperature, a single peak starts to broaden and split into double peaks at around $T$*=215 K, as can be clearly seen in Fig. 6. This temperature $T$* is 35 K higher than the macroscopic phase transition temperature $T_{d1}$=180 K determined by elastic measurements. This means that in this structural phase transition, the critical slowing down takes place ahead of the divergence of the correlation length, bringing the system to a spatially-incoherent and static phase in rather a wide temperature region above the critical temperature of the macroscopic phase transition. Such an anomalous behaviour was first reported by Fukase *et al.* in the midst of 80's as the anomalous reduction in elastic constant from $T$* which is much higher than $T_{d1}$.[6] The determination of $T_{d1}$ by the elastic measurements is shown in Appendix [15,17].

Finally, we refer to the inequivalent Cu sites in La-based cuprates. So far, many NQR reports confirm the existence of the inequivalent Cu sites with different EFG amplitude[21]. It was called as B-site against the main A-site. Its number fraction is nearly 10 % of the main A-site for the Sr concentration $x$=0.15, and increases with Sr-concentration. In our rotation or field-swept spectra, however, we did not observe the B-site within experimental precision, though we did in zero-field NQR measurements [21,22]. This fact shows that the EFG principal axis of B-site is not unique and along the $c$-axis, but includes at least four directions, reflecting the local symmetry. Therefore, resonance lines split in measurements under a finite field, and its amplitude is reduced to be far behind the detection limit.

## Summary

We have investigated the local structure in single crystals of LSCO with $x$=0.12 and 0.15 by the Cu-NMR spectra. In $x$=0.12, we have demonstrated an appearance of spatially incoherent LTT phase below 50 K as the change in the spectral profile. This LTT phase is considered to be the origin of the weak $T_C$ anomaly around $x$=1/8 in rare-earth free LSCO. In $x$=0.15, we have investigated the structural phase transition from HTT to LTO and found that the change in the local buckling pattern starts much higher than the macroscopic structural phase transition temperature.

## Acknowledgments

The authors are grateful to Dr. T. Hanaguri for his kind permission for using the data of elastic measurements and X-ray powder diffraction at low temperatures. This work was supported by a Grant-in-Aid for Scientific Research on the Priority Area “High Magnetic Field Spin Science in 100T ” from MEXT.

## Appendix

The structural phase transition from HTT phase to LTO is sensitively detected by elastic measurement, and its transition temperature $T_{d1}$ is determined from the temperature dependence of the ultrasonic sound velocity as the crossing point of linear extrapolations from the both sides of the transition. Thus determined transition temperature precisely agrees with the macroscopic structural phase transition, that is, the onset of the orthorhombicity observed by X-ray diffraction as seen in Fig. A1 [6,15]. The transition temperature $T_{d2}$ can also be determined from the ultrasonic velocity as a slight hardening point.

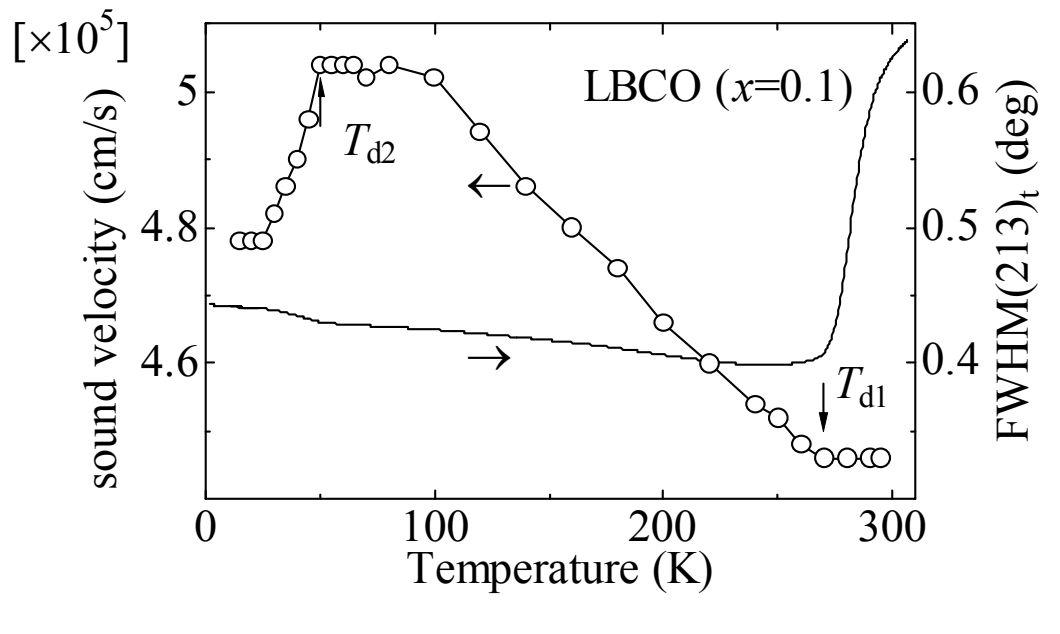

**Figure A1.** Temperature dependence of ultrasonic sound velocity measured by longitudinal wave of 7 MHz and the orthorhombic splitting of Bragg peak $(213)_t$ in polycrystalline sample of LBCO ($x$=0.1). The critical temperature of the structural phase transition from HTT to LTO, $T_{d1}$, and from LTO to LTT, $T_{d2}$ are shown by arrows.